\documentclass[twocolumn,tighten]{aastex62}

\received{May 10, 2019}
\revised{July 31, 2019}
\accepted{August 9, 2019}
\submitjournal{ApJ}

\usepackage{url, soul, bm}
\usepackage{amsmath}      

\shorttitle{X-ray Emission from FU Ori-type Stars}
\shortauthors{Kuhn et al.}

\begin{document} 

\title{A Comparison of the X-ray Properties of FU Ori-type Stars to Generic Young Stellar Objects}

\correspondingauthor{Michael A. Kuhn}
\email{mkuhn@astro.caltech.edu}

\author[0000-0002-0631-7514]{Michael A. Kuhn}
\affil{Department of Astronomy, California Institute of Technology, Pasadena, CA 91125, USA}

\author{Lynne A. Hillenbrand}
\affil{Department of Astronomy, California Institute of Technology, Pasadena, CA 91125, USA}

\begin{abstract} 
Like other young stellar objects (YSOs), FU Ori-type stars have been detected as strong X-ray emitters. However, little is known about how the outbursts of these stars affect their X-ray properties. We assemble available X-ray data from {\it XMM Newton} and {\it Chandra} observations of 16 FU~Ori stars, including a new {\it XMM Newton} observation of Gaia~17bpi during its optical rise phase. Of these stars, six were detected at least once, while 10 were non-detections, for which we calculate upper limits on intrinsic X-ray luminosity ($L_X$) as a function of plasma temperature ($kT$) and column density ($N_H$). The detected FU~Ori stars tend to be more X-ray luminous than typical for non-outbursting YSOs, based on comparison to a sample of low-mass stars in the Orion Nebula Cluster. FU~Ori stars with high $L_X$ have been observed both at the onset of their outbursts and decades later. We use the Kaplan-Meier estimator to investigate whether the higher X-ray luminosities for FU~Ori stars is characteristic or a result of selection effects, and we find the difference to be statistically significant ($p<0.01$) even when non-detections are taken into account. The additional X-ray luminosity of FU~Ori stars relative to non-outbursting YSOs cannot be explained by accretion shocks, given the high observed plasma temperatures. This suggests that, for many FU~Ori stars, either 1) the outburst leads to a restructuring of the magnetosphere in a way that enhances X-ray emission, or 2) FU~Ori outbursts are more likely to occur among YSOs with the highest quiescent X-ray luminosity.
\end{abstract}

\keywords{
accretion disks, stars: formation, stars: individual (Gaia 17bpi), stars: variables: T Tauri, Herbig Ae/Be, X-ray: stars}

\section{Introduction}\label{intro.sec}

An FU Ori-type star is a young stellar object (YSO) in a state of outburst, during which its optical brightness increases by 4--6~mag on a timescale of months to years and declines over a time scale of decades to centuries \citep{2010vaoa.conf...19R}. These events are caused by a change in structure of the circumstellar disk, resulting in high accretion rates and causing the disk to outshine the central star in the optical and infrared \citep{1996ARA&A..34..207H}. However, our understanding of these objects is based on only a handful of sources where the initial outburst was observed, along with a small number of other stars that exhibit FU Ori-like characteristics \citep[see][]{2018ApJ...861..145C}.

The changes that produce these outbursts could also be expected to affect the X-ray emission from these YSOs. For ordinary pre--main-sequence stars, X-ray emission is dominated by coronal plasmas with temperatures of tens of millions of degrees Kelvin heated by magnetic reconnection \citep{1999ARA&A..37..363F,2004A&ARv..12...71G}. This mechanism is thought to be similar to the process that heats the corona of main-sequence stars like the Sun, but is scaled up by a factor of $10^3$--$10^{5}$ due to stronger magnetic dynamos in pre--main-sequence stars \citep{2005ApJS..160..401P}. Accretion onto YSOs can also produce shock heating. The maximum temperature of gas heated this way is limited to several million Kelvin by the free-fall velocity onto the star, yielding a cooler component of X-ray emission that is more easily absorbed by surrounding material. However, accretion generated X-ray emission has been observed in some T Tauri stars with X-ray spectroscopy \citep[][]{2002ApJ...567..434K,2005A&A...432L..35S}. Outflowing jets are another possible source of X-rays from YSOs. X-ray observations of these jets have typically revealed shock-heated gas with temperatures of several million Kelvin \citep{2007A&A...462..645B}.

Several processes in FU Ori stars could either suppress or enhance the X-ray production mechanisms described above. 
During an outburst it is thought that the circumstellar disk moves inward toward the star, possibly crushing the stellar magnetosphere \citep{1994ApJ...429..781S}. However, there has been little examination of how this would affect X-ray production. 
For non-outbursting T Tauri stars, observations have indicated that accretion suppresses X-ray emission \citep{2007A&A...468..425T}, possibly as the result of mixing between accreted gas and magnetically heated plasma, cooling the hot gas. It also seems likely that changes in the configuration of material in the disk, accretion streams, or outflows could increase the absorption of X-ray emission produced close to the central star. On the other hand, higher accretion rates and stronger jets seem likely to enhance soft X-ray emission.

Our understanding of X-ray emission from FU Ori stars has been based on a small number of studies of individual objects, including FU Ori itself \citep{2010ApJ...722.1654S}, \object[HBC 722]{HBC~722} \citep{2014A&A...570L..11L}, \object[V960 Mon]{V960~Mon} \citep{2015ATel.7025....1P}, \object[V1735 Cyg]{V1735~Cyg} \citep{2009ApJ...696..766S}, \object[L1551 IRS 5]{L1551 IRS 5} \citep{2003ApJ...584..843B,2011A&A...530A.123S}, and \object[Z CMa]{Z CMa} \citep{2009A&A...499..529S}. Several of these studies have noted particularly high X-ray luminosities compared to what is expected for typical YSOs \citep[e.g.,][]{2009ApJ...696..766S,2010ApJ...722.1654S}. Furthermore, \citet{2014A&A...570L..11L} reported increasing X-ray luminosity of HBC~722 in three observations during the first several years of its outburst. The X-ray emission from most of these objects is dominated by a high temperature component that is most likely the result of magnetic reconnection heating. However, non-detections, which have resulted from most X-ray observations, have been largely ignored\footnote{Upper limits have been published for V1057 Cyg and V1515 Cyg by \citet{2010ApJ...722.1654S} and a non-detection in an initial observation of HBC~722 is discussed by \citet{2014A&A...570L..11L}. Additional detections and/or non-detections are available in archival datasets for those and other FU Ori stars as presented in Section~\ref{constraints.sec}.}, which results in a selection bias in the literature that makes it difficult to draw conclusions about the overall population. 

In order to better understand possible changes in X-ray properties of FU Ori stars, we acquired an {\it XMM Newton} observation of the most recent FU Ori star, \object[Gaia 17bpi]{Gaia~17bpi}, shortly after its discovery. Gaia~17bpi is a spectroscopically confirmed FU Ori-type star \citep{2018ApJ...869..146H} that triggered a Gaia Science Alert \citep{2013RSPTA.37120239H} in 2017. The source has continued to brighten in the {\it Gaia} $G$ band, increasing by 4~mag as of mid-2019.

We pool the available X-ray observations of FU Ori stars, including Gaia~17bpi, to investigate whether there is sufficient evidence to determine whether the outbursts affect X-ray emission from these stars. Section~\ref{data.sec} describes the available X-ray data for FU Ori stars and our reduction methodology for Gaia~17bpi. In Section~\ref{constraints.sec} we derive constraints on X-ray fluxes and model parameters for non-detected FU Ori stars and compare these to expectations for non-outbursting YSOs. Section~\ref{other.sec} examines the properties of the full FU Ori sample, constructing an X-ray luminosity function that includes both detections and non-detections and comparing this distribution to the X-ray luminosity function for ordinary YSOs. Finally, Section~\ref{discussion.sec} discusses implications of enhanced X-ray emission observed from some FU Ori stars.

\section{X-ray Observations and Data Reduction}\label{data.sec}

\subsection{{\it XMM Newton} observation of Gaia 17bpi}\label{gaia17bpi_reduction.sec}
Gaia~17bpi was observed by {\it XMM Newton} at our request as an unanticipated target of opportunity (TOO) after its classification as an FU Ori type star. The observation (ID~0821400201; PI: Norbert Schartel) was made during {\it XMM Newton} revolution 3468 on November 15, 2018 with a nominal aim-point of 19:31:05.60 +18:27:52.0. Observations with EPIC's MOS1 (36747~s), MOS2 (36731~s), and pn (35203~s) detectors were made in full frame mode using the thin filter. RGS1 and RGS2 were also active during the observation; however, the low flux from Gaia~17bpi means that only the EPIC instrument provides constraints on X-ray flux. The optical monitor (OM) took a series of exposures, rotating sequentially through the $V$, $U$, and $UVW1$ filters.

X-ray imaging array data from EPIC were reduced using the {\it XMM Newton} threads with SAS version 17.0.0 (SAS Development Team 2014), HEASOFT version 6.25 (HEASARC 2014), and calibration files from November 27, 2018. We created calibrated event files using the SAS procedures {\it epproc} and {\it emproc}, removed time intervals with background flaring, and applied a barycentric correction. 
We retain events from the pn and MOS detectors with $\mathtt{PATTERN}<12$, $0.2<\mathtt{PI}<12$~keV, and the standard {\tt \#XMMEA\_EP} and {\tt \#XMMEA\_EM} flags.  

An X-ray source was not apparent at the location of Gaia~17bpi in the EPIC images. Our analysis focuses on the pn detector due to its greater sensitivity.  To determine limits on X-ray flux from Gaia~17bpi, source extraction was performed using a circular aperture at the location of the star (19:31:05.60 +18:27:52.0 ICRS) with a radius of 15$^{\prime\prime}$. We also defined 9 background apertures of the same size on the same chip with distances from the target source ranging from 1$^\prime$ to 2$^\prime$, which were selected to avoid other nearby X-ray sources. Observational properties that affect the conversion between detector events and photons, including telescope throughput, detector sensitivity, vignetting, bad pixels, and light lost from the aperture are encoded in the auxiliary response file (ARF) generated by {\it arfgen}.

The OM images were reduced by the Processing Pipeline Subsystem (PPS) to yield a ultraviolet/optical catalog. Gaia~17bpi was detected in the $U$ and $V$ bands, but not in $UVW1$, by the {\it omdetect} algorithm, and the source was extracted using 6-pixel apertures. The photometry in the XMM-Newton OM Vega-magnitude system is $UVW1>20.4$~mag, $U=21.12\pm0.23$~mag, and $V=17.75\pm0.04$~mag.  

\subsection{X-ray Data for other FU Ori Stars}\label{data_other.sec}

We compare X-ray constraints on Gaia~17bpi to properties of other FU Ori stars that were previously observed by {\it XMM Newton} or {\it Chandra}. We base this comparison on the list of FU Ori-type stars compiled by \citet{2018ApJ...861..145C}, which includes both {\it bona fide} outbursts as well as the FU Ori-like sources that have spectroscopic features of FU Ori stars but did not have observed eruptions.\footnote{Our final list includes 13 objects from the ``{\it bona fide}'' class as well as 3 FU Ori-like objects (L1551 IRS~5, IRAS 05450+0019, and Z CMa). We use both classes in this paper.} Despite the passage of multiple decades since the outbursts of some of these objects, they are still elevated in brightness compared to their pre-outburst level, albeit some have decayed significantly \citep[e.g., V1057 Cyg;][]{2005MNRAS.361..942C}. Furthermore, all the stars in the sample continue to have spectroscopic characteristics of FU Ori stars \citep{2018ApJ...861..145C}. 

Our sample of FU Ori stars includes all stars from the above list for which {\it Chandra} or {\it XMM Newton} data is publicly available as of May 2019. In addition to the published X-ray sources listed in Section~\ref{intro.sec}, we also include unpublished archival X-ray data for V883 Ori, V2775 Ori, V900 Mon, V2494 Cyg, V2495 Cyg, V733 Cep, V1057 Cyg, V1515 Cyg, and IRAS 05450+0019, and V960~Mon (second epoch).

Depending on the availability of data for each FU Ori star, several methods have been used to assemble X-ray properties. For cases where model fits to the X-ray data have been published, we use these models to calculate the absorption-corrected X-ray luminosities $L_X$ in the 0.5--8.0~keV band -- i.e.\ the integrated X-ray luminosity of the model with absorption set to zero. In other cases we make use of publicly available catalogs, such as the XMM Newton Serendipitous Source Catalog data release 7 \citep[3XMM-DR7;][]{2019yCat.9054....0R} and the Star Formation in Nearby Clouds catalog \citep[SFiNCs;][]{2017ApJS..229...28G}, as well as data from the {\it XMM Newton} and {\it Chandra} archives, which we reduce using the standard recipes (Appendix~\ref{individual.sec}).

\section{Constraints on X-ray Properties for Undetected Sources}\label{constraints.sec}

\subsection{Upper Limit on X-ray Count Rate}\label{upperlimit.sec}

Most of the non-detections of FU Ori stars come from {\it XMM Newton} observations (13 out of 14). Eleven out of the 13 are located in regions of the sky covered by 3XMM-DR7. For these objects, the FLIX tool\footnote{\url{https://www.ledas.ac.uk/flix/flix_dr7.html}} \citep[][their Appendix~A]{2007A&A...469...27C} is used to calculate upper limits on flux in the 5 {\it XMM Newton} energy bands (Table~\ref{limits.tab}) at each stars' position based on the X-ray images that were used to generate the 3XMM-DR7 catalog. 

The other two {\it XMM Newton} non-detections include our observation of Gaia~17bpi and a recent observation of V900~Mon, which we analyze using the same method as detailed below for Gaia~17bpi.   The final non-detection is a Chandra observation of V733~Cep, for which we use the completeness limit of the SFiNCs catalog. Details of the process for individual FU Ori stars are further described in Appendix~\ref{appendix_nondetections.sec}.
 
We compute upper limits on the flux of Gaia~17bpi in each of the bands listed above using the data from the pn detector. In each band, the upper limits for the number of X-ray events from Gaia~17bpi ($N_\star$) are determined by counting the number of events in the source extraction region ($N_\mathrm{ext}$) and comparing this to numbers of events in background regions ($N_\mathrm{bgd}$). We assume that the number of events in each extraction region is drawn from a Poisson distribution, and we find that the variations in numbers of counts in each of the nine background regions are consistent with this assumption. Then, the likelihood of $N_\star$ given that $N_\mathrm{ext}$ counts were observed would be the Poisson probability
\begin{equation}
p(N_\mathrm{ext}|N_\star) = \frac{(N_\star + N_\mathrm{bgd})^{N_\mathrm{ext}}e^{-(N_\star + N_\mathrm{bgd})}}{N_\mathrm{ext}!}.
\end{equation}

\begin{deluxetable}{lr|rrr}[t]
\tablecaption{X-ray Flux Upper Limits for Gaia~17bpi\label{limits.tab}}
\tabletypesize{\small}\tablewidth{0pt}%\rotate
\tablehead{
  \multicolumn{2}{c}{Band} &    \multicolumn{3}{c}{Flux} \\
  \colhead{$E_\mathrm{min}$} &  \colhead{$E_\mathrm{max}$}   &  \colhead{95\%} &\colhead{99\%} & \colhead{99.9\%} \\
 \colhead{keV} &  \colhead{keV}  & \multicolumn{3}{c}{erg~s$^{-1}$\,cm$^{-2}$}
}
\startdata
 0.2&0.5 & $2.7\times10^{-16}$ & $3.8\times10^{-16}$  & $5.3\times10^{-16}$  \\
 0.5&1.0 & $3\times10^{-16}$ &  $5\times10^{-16}$ & $7\times10^{-16}$  \\
 1.0&2.0 & $6\times10^{-16}$ &  $8\times10^{-16}$ & $1.1\times10^{-15}$  \\
 2.0&4.5 & $1.7\times10^{-15}$ &  $3\times10^{-15}$ & $4\times10^{-15}$ \\
 4.5&12.0 & $8\times10^{-15}$ &  $1.1\times10^{-14}$ & $1.6\times10^{-14}$ \\
\enddata
\end{deluxetable}

As few FU Ori type stars have been detected in the X-ray, the existing fluxes provide little guidance for what to use for a prior. For pre--main-sequence stars, studies of the Orion Nebula Cluster reveal a distribution that is relatively constant in log X-ray luminosity ($\log L_X$) over the range from $10^{28}$~erg~s$^{-1}$ to $2\times10^{30}$~erg~s$^{-1}$,  and decreases for higher luminosities \citep[][]{2005ApJS..160..379F}.  For our prior distribution on the flux of Gaia~17bpi, we used a uniform distribution in $\log F_X$ between $-17.0$ and $-12.5$, which spans the full range of plausible upper limits. However, we found that if we use an alternative prior, generated by adaptively smoothing the distribution of X-ray fluxes of stars in the Orion sample, the results are very similar.

From Bayes' theorem we know,
\begin{equation}
p(N_\star) = \frac{p(N_\mathrm{ext}|N_\star)\,p(N_\star)}{\int_{0}^{\infty}p(N_\mathrm{ext}|N_\star)\,p(N_\star)\mathrm{d}N_\star}.
\end{equation}
The integral in the denominator can be computed numerically, allowing us to compute the credible interval on the range $[0,N_{\star,ul}]$ that contains 99\% of the probability, where $N_{\star,ul}$ is the 99\% upper limit. We also compute 95\% and 99.9\% upper limits.

The upper limits on $N_\star$ are converted to count rate by dividing by $\mathtt{ontime}$ (32,694~s at the position of Gaia~17bpi). Count rate is converted to flux by dividing by the mean value of the ARF in the band and multiplying by the central energy \citep[see approximation by][their Equation~8]{2010ApJ...714.1582B}. The 95\%, 99\%, and 99.9\% upper limits on flux are reported in Table~\ref{limits.tab}. Flux limits are less restrictive for the higher energy bands because the pn detector is less sensitive to photons in these energy ranges and the flux would be distributed among fewer photons. 

\subsection{Constraints on $L_X$, $kT$, and $N_H$}\label{lum.sec}

\begin{figure*}[t]
\centering
\includegraphics[width=0.75\textwidth]{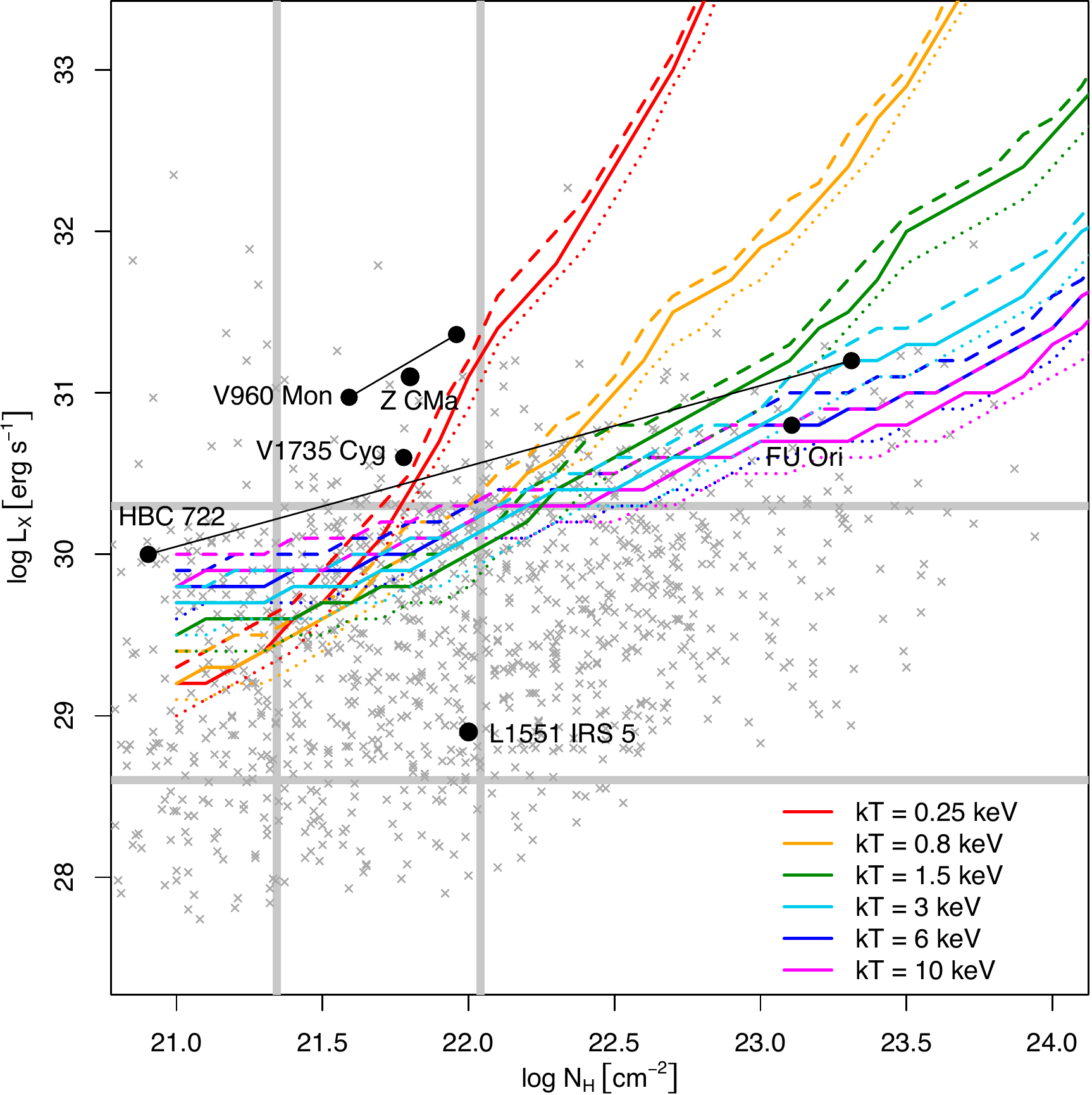} 
\caption{The colored lines show the maximum possible $L_X$ for Gaia~17bpi as a function of $N_H$ for various assumptions of plasma temperature -- red: $kT=0.25$~keV, orange: 0.8~keV, green: 1.5~keV, cyan: 3~keV, blue: 6~keV, and magenta: 10~keV. The dotted, solid, and dashed lines indicated the 95\%, 99\%, and 99.9\% probability limits, respectively. The light gray vertical lines indicate the range of $N_H$ that would correspond to the pre-burst $A_V=1$--5~mag estimated by \citet{2018ApJ...869..146H}, while the horizontal lines show the probable range of pre-burst $L_X$ estimated from the $L_X$--$L_{\mathrm{bol}}$ relation. The gray X's show $L_X$ and $N_H$ for low-mass, non-flaring pre--main-sequence stars in the Orion Nebula Cluster \citep{2005ApJS..160..319G}, while black points indicated previously detected FU Ori-type stars. Black lines connect points for FU Ori stars observed at different epochs. \label{main.fig}}
\end{figure*}

The upper limits on X-ray fluxes can be used to determine what combinations of X-ray luminosity, plasma temperature, and absorbing column are possible. For the non-detections, we calculate upper limits on X-ray luminosity as a function of $N_H$ and $kT$. The procedure is described in detail below for Gaia~17bpi and this case is illustrated in Figure~\ref{main.fig}.
The limits for other sources are summarized in Table~\ref{other.tab}.

We use the Portable, Interactive Multi-Mission Simulator \citep[PIMMS;][]{1993Legac...3...21M} to calculate X-ray flux in the five {\it XMM Newton} bands for a grid of temperatures ($kT=0.25$, 0.8, 1.5, 3.0, 6.0, and 10.0~keV), absorptions ($N_H=10^{21}$ to $10^{24}$~cm$^{-2}$), and intrinsic (i.e.\ unabsorbed) X-ray luminosities ($L_X=10^{28}$ to $10^{33}$~erg~s$^{-1}$ in the 0.5--8.0~keV band), given the distances to the FU Ori stars from Appendix~\ref{distances.sec}. For these simulations we set metal abundances of the model to 0.4 times solar, which are typical for X-ray emission from YSOs \citep[e.g.,][]{2001ApJ...557..747I,2002ApJ...574..258F}; however, this assumption has little effect on the resulting limits. A combination of $L_X$, $kT$, and $N_H$ is ruled out if it produces a flux in one of the five bands that exceeds the flux limits from Table~\ref{limits.tab}. 

The colored lines on Figure~\ref{main.fig} show limits on $L_X$ for Gaia~17bpi as a function of $N_H$. Each color shows limits for a different temperature plasma, while the dotted, solid, and dashed lines refer to the 95\%, 99\%, and 99.9\% limits. Other symbols and annotation on the graph are explained in Sections~\ref{pre.sec}--\ref{other.sec} below. 

First, we examine whether a lower temperature plasma ($kT\sim0.25$~keV, as would be expected for accretion shocks or jets) would have been detectible. If the source were lightly absorbed ($N_H<10^{21.6}$~cm$^{-2}$), it must be less luminous than $\sim$10$^{29}$--$10^{30}$~erg~s$^{-1}$. However, for somewhat higher absorption, there is little constraint on the possible luminosity of a soft X-ray source. 

For higher temperature plasmas ($kT\sim0.8$--$3.0$~keV, typical for coronal X-ray emission from T Tauri stars), X-ray luminosities above $\sim$10$^{30}$~erg~s$^{-1}$ can be excluded if the absorption is not much greater than $N_H=10^{22}$~cm$^{-2}$, as inferred from the optical pre-outburst observations (Section~\ref{pre.sec}). However, if the post-outburst absorption were greater, the range of possible X-ray luminosities increases with increasing $N_H$.

For very hot plasmas ($kT>10$~keV, similar the the hottest X-ray emission from T Tauri stars or X-ray emission from some FU Ori stars), possible X-ray luminosities range from $<$10$^{30}$~erg~s$^{-1}$ for light absorption to $<$10$^{30.8}$~erg~s$^{-1}$ for the heavy extinction of $N_H\sim10^{23.5}$~cm$^{-2}$ (equivalent to $A_V\sim140$~mag with a typical gas-to-dust law).

The curves of $L_X$ upper limits for the other FU Ori stars with non-detections look similar to those of Gaia~17bpi, with vertical shifts resulting from different distances and integration times. In order to compare upper limits for different objects, it is helpful to pick a fiducial ($N_H$, $kT$) combination. We choose $kT=3$~keV and $N_H=10^{21.8}$~cm$^{-2}$, and we tabulate the upper limits on $L_X$ in Table~\ref{other.tab} using these assumptions. These temperatures and absorptions are close to the median values obtained from the sample of 6 FU Ori stars where these quantities have been measured (also shown in Table~\ref{other.tab}). However, it is possible that the upper limits on $L_X$ in Table~\ref{other.tab} would be too low if $N_H$ were much greater than $N_H=10^{21.8}$~cm$^{-2}$ -- implications of the ($N_H$, $kT$) choice are discussed in Section~\ref{discussion.sec}. 

\subsection{Expected Ranges of Pre-Outburst $N_H$ and $L_X$ for Gaia~17bpi}\label{pre.sec}

Gaia~17bpi was not observed in the X-ray prior to the outburst. However, probable ranges for pre-outburst $N_H$ and $L_X$ can be estimated from statistical relations between X-ray properties and stellar properties for T~Tauri stars. The pre-outburst spectral energy distribution (SED) of Gaia~17bpi could be fit by a variety of reddened stellar photospheric models with $A_V$ ranging from 1 to 5~mag and corresponding $T_\mathrm{eff}$ ranging from 2900 to 4500~K \citep{2018ApJ...869..146H}, with a preferred model with $T_\mathrm{eff}=3500$~K and $A_V=3$~mag (see also Appendix~\ref{om.sec}). From these ranges, we derive extinction-corrected bolometric luminosities of $L_{bol} \approx 4\times10^{32}$--$2\times10^{33}$~erg~s$^{-1}$ for the pre-outburst star using bolometric and color corrections from \citet{2013ApJS..208....9P}.

\citet{1996Ap&SS.236..285R} found $N_H/A_V\approx2.2\times10^{21}~\mathrm{cm}^{-2}/\mathrm{mag}$ for Galactic extinction, and this law provides a reasonable description for the observed relation between $A_V$ and $N_H$ for T Tauri stars \citep[e.g.,][]{2005ApJS..160..379F,2010ApJ...725.2485K}. For Gaia~17bpi, this would yield a pre-burst $N_H=2.2\times10^{21}$--$1.1\times10^{22}$~cm$^{-2}$. This range is indicated in Figure~\ref{main.fig} for comparison. 

X-ray luminosity is correlated with bolometric luminosity (or mass) for T Tauri stars, 
 albeit with significant scatter. 
 From the $L_X$--$L_{bol}$ relation\footnote{We use the regression line and standard deviation derived for ``all stars'' in Table~4 of \citet{2016MNRAS.457.3836G}.} in \citet[][]{2016MNRAS.457.3836G}, we would expect a pre-outburst X-ray luminosity to fall in the range
 $L_X=4\times10^{28}$--$2\times10^{30}$~erg~s$^{-1}$ taking into account both uncertainty on the model and statistical scatter. 
 Given that this range spans the upper limit on $L_X$ of Gaia~17bpi as indicated on Figure~\ref{main.fig}, our constraints do not reveal whether X-ray flux was suppressed by the beginning of the outburst relative to the expected pre-outburst emission. 

\subsection{Comparison with Generic YSOs}\label{generic_yso.sec}

We use a sample of pre--main-sequence stars detected in the Chandra Orion Ultradeep Project \citep[COUP;][]{2005ApJS..160..319G,2005ApJS..160..379F} as a representative sample of generic YSOs. We include only cluster members that are low-mass (excluding stars with spectral-types of B and earlier), that do not have major X-ray flares during the observation \citep[the list of flaring sources is taken from][]{2008ApJ...688..418G}, and have measurements of $L_X$ and $N_H$. This yields a sample of 1200 objects that we refer to as the ``COUP sample.'' 

The COUP sample is plotted on Figure~\ref{main.fig}. These stars (gray x's) mostly lie within the same region of the plot as the permitted Gaia~17bpi X-ray luminosities; however, some of the most luminous COUP sample X-ray sources would have been detected in this observation. This suggests that if Gaia~17bpi did have X-ray properties that were unchanged by the outburst, it would most likely not have been detected.

\section{X-ray Properties of the Full FU Ori Sample}\label{other.sec}

In our full sample, six FU Ori stars were detected in the X-ray, while ten were not (Table~\ref{other.tab}). Several stars were observed at multiple epochs during the outburst or shortly afterwards, and these observations are listed separately in the table in order to document any evolution of X-ray properties during an outburst (see also Appendix~\ref{individual.sec}). The detected FU Ori stars have quiescent X-ray light curves, with the possible exception of the 2008 {\it Chandra} epoch of FU Ori, where X-ray variability is interpreted by \citet{2010ApJ...722.1654S} as a possible flare. 

The six detected FU Ori-type stars are plotted on Figure~\ref{main.fig} (black points).  Five of these lie above the $L_X$ curve for Gaia~17bpi for the corresponding $kT$ on this figure, indicating that they would have been detected if they were in this observation at the same distance as Gaia~17bpi. This implies that Gaia~17bpi is currently fainter in the X-ray than these 5 stars. These five stars are also much brighter than the typical YSOs in the COUP sample in the Orion Nebula Cluster, meaning that they either represent the high-luminosity tail of the pre--main-sequence star X-ray luminosity function, or their X-ray luminosities have been enhanced by their outbursts. One object, L1551 IRS 5, has an $L_X$ much lower than the other detected FU Ori stars.

\begin{deluxetable*}{lrccclcrr}[t]
\tablecaption{Summary of X-ray properties of FU Ori stars\label{other.tab}}
\tabletypesize{\small}\tablewidth{0pt}
\tablehead{
  \colhead{Name} &  \colhead{$\log L_X$\tablenotemark{a}}   &  \colhead{$kT$\tablenotemark{b}} &\colhead{$\log N_H$} & \colhead{Facility} & \colhead{Date\tablenotemark{c}} & \colhead{X-ray Ref.\tablenotemark{d}} & \colhead{Region}  & \colhead{Dist.\tablenotemark{e}}\\
  \colhead{} &  \colhead{erg s$^{-1}$}   &  \colhead{keV} &\colhead{cm$^{-2}$}  & \colhead{} & \colhead{UT} & \colhead{} & \colhead{}  & \colhead{pc}
}
\startdata
Gaia 17bpi & $<$29.9  & $[3.0]$ & $[21.8]$   & XMM & 2018-11-15 & this work & G53.2 & 1200\\
Z CMa & 31.1  & 0.4+7.5 & 21.8   & CXO & 2003-12-07 & Stelzer et al. (2009) & CMa R1&  1120\\ 
V900 Monocerotis & $<$29.8  & $[3.0]$ & $[21.8]$   &  XMM & 2016-10-10 & this work & CMa R1 & 1120\\ 
V960 Monocerotis & 31.0 & 2.6 & 21.6  & CXO & 2015-01-26 & this work & CMa R1 &   1120\\
~~~~---&   31.4  & 0.5+2.6  & 22.0  &   XMM & 2016-10-28 & this work & ---&--- \\
V1515 Cygni & $<$29.9 & $[3.0]$ & $[21.8]$ & XMM & 2006-10-22 & 3XMM-DR7 & NGC 6914&  960 \\
V733 Cephei & $<$30.0  &\nodata  & $[21.8]$   & CXO &  2009-04-28  & this work & Cep OB3& 825 \\ 
~~~~--- & $<$30.6  & $[3.0]$ & $[21.8]$ & XMM & 2014-02-28  & 3XMM-DR7 & ---& ---\\ 
V1057 Cygni & $<$29.9  & $[3.0]$ & $[21.8]$   & XMM & 2005-11-26  & 3XMM-DR7 & NGC 7000 &  795\\ 
HBC 722 & $<$30.8  & $[3.0]$ & $[21.8]$   & XMM & 2010-11-25  & 3XMM-DR7 & NGC 7000 &  795\\
~~~~--- & 30.0  & 8.4 & 20.9 &   XMM & 2011-05-26 & Liebhart et al.\ (2014) &---& ---\\
~~~~--- & 31.2  & 2.3 & 23.3 &  CXO & 2013-07-17 &Liebhart et al.\ (2014) &---& ---\\ 
V1735 Cygni & 30.6  & 13.7 & 21.8  & XMM & 2006-07-22 & Skinner et al.\ (2009) & IC 5146 &  752\\
V2494 Cygni & $<$29.6  &  $[3.0]$ & $[21.8]$   & XMM & 2013-12-06 & 3XMM-DR7 & Cyg OB7 &  594\\
V2495 Cygni & $<$29.4  &  $[3.0]$ & $[21.8]$   & XMM & 2013-04-19 & 3XMM-DR7 & Cyg OB7&  594\\
~~~~---   &   $<$29.5  &  $[3.0]$ & $[21.8]$   & XMM & 2013-12-05 & 3XMM-DR7 & ---&  ---\\
IRAS 05450+0019  & $<$29.4 & $[3.0]$ & $[21.8]$  & XMM & 2005-03-30 & 3XMM-DR7& NGC 2071 &  446\\ 
V2775 Orionis & $<$30.0  & $[3.0]$ & $[21.8]$   & XMM & 2007-09-20 & 3XMM-DR7& Orion A & 416\\
FU Orionis & 30.8  & 5.58 & 23.1   & XMM & 2004-03-08 & Skinner et al.\ (2006) & $\lambda$ Ori & 404 \\ 
~~~~--- & 30.8  & 4.50 & 23.1   & CXO & 2008-11-24 & Skinner et al.\ (2010) & --- & --- \\ 
V883 Orionis & $<$29.7  & $[3.0]$ & $[21.8]$  & XMM & 2004-03-27 & 3XMM-DR7& Orion A & 392 \\
~~~--- & $<$29.6  & $[3.0]$ & $[21.8]$ & XMM & 2004-08-25 & 3XMM-DR7& ---  & ---  \\
L1551 IRS 5 & 28.9 & 0.6& 22.0  & XMM+CXO& 2000-09-09 & Schneider et al. (2011) & L1551 & 144\\
\enddata
\tablecomments{Stars are sorted in inverse order of distance.}
\tablenotetext{a}{Absorption-corrected X-ray luminosities. All values are adjusted for assumed distance and to a common X-ray band (0.5--8.0~keV).} 
\tablenotetext{b}{Brackets around values indicate that they are assumed, not derived from X-ray spectra.} 
\tablenotetext{c}{Start time of the first observation used for the analysis.}
\tablenotetext{d}{Reference to the paper that provides the X-ray model or the catalog that provides the flux upper limits. Objects for which X-ray properties are newly derived are labeled ``this work.''}
\tablenotetext{e}{See Appendix~\ref{distances.sec} for justification of adopted distances, which includes original work.}
\end{deluxetable*}

\subsection{Time Evolution of X-ray Luminosity}\label{evolution.sec}

\begin{figure}[t]
\centering
\includegraphics[width=0.48\textwidth]{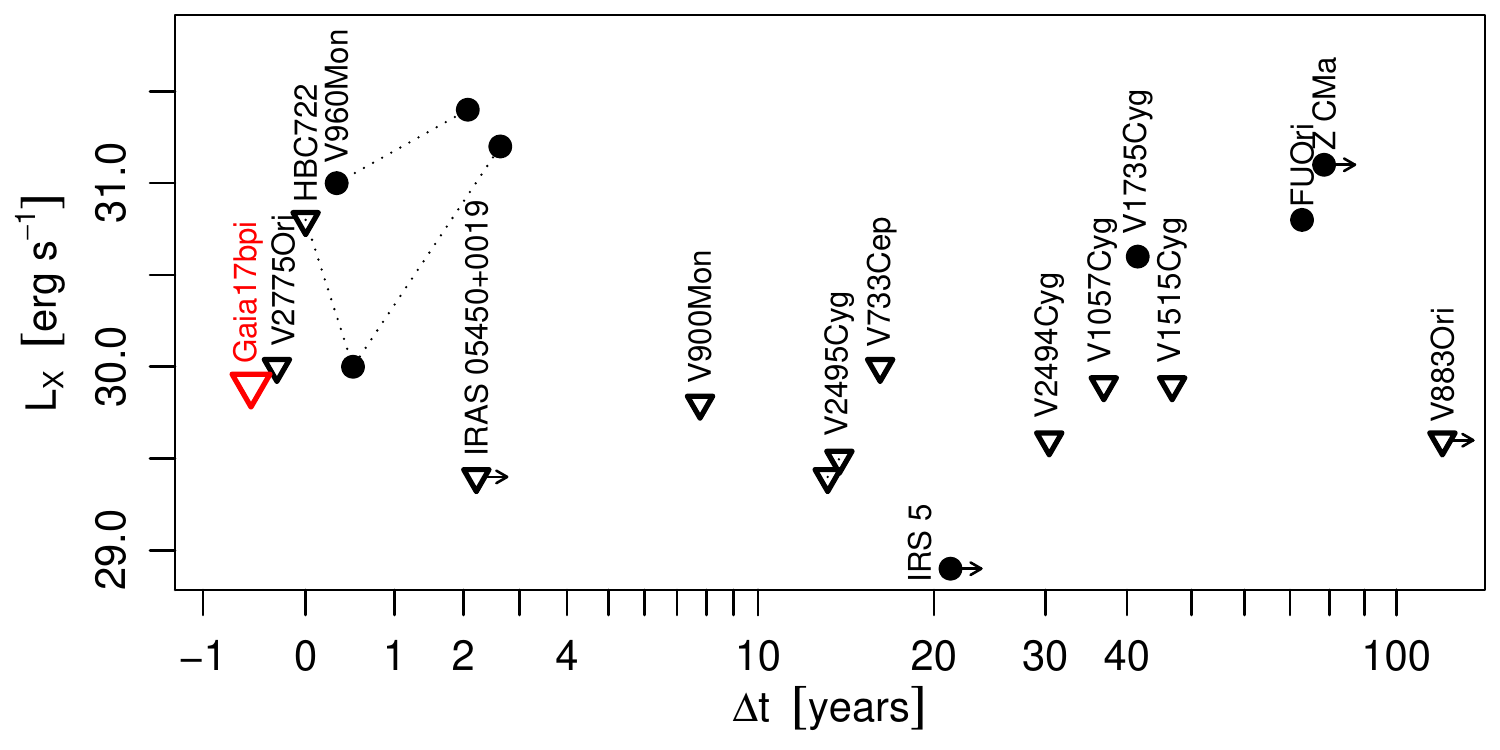} 
\caption{X-ray luminosities of FU Ori type stars observed with {\it Chandra} or {\it XMM Newton} plotted as a function to the time $\Delta t$ between the peak of the optical outburst and X-ray observation. Detections are shown as filled circles and non-detections as downward-pointing triangles. Arrows indicate that time since outburst is a lower limit for four objects. Objects that were observed at multiple epochs are joined with dashed lines. The $\Delta t$ axis is approximately linear for $\Delta t \approx 0$ and becomes logarithmic for larger $\Delta t$ due to the large dynamic range in time separation.
 \label{timeline.fig}}
\end{figure}

Figure~\ref{timeline.fig} shows X-ray luminosities (or upper limits on X-ray luminosity) for the FU Ori stars as a function of the time elapsed between the peak of the optical outburst and the X-ray observation.  The dates of the outburst peaks are taken from \citet{2018ApJ...861..145C}; however, in many cases these are not well constrained due to either the cadence of observing, ambiguity in the shape of the light curve, or (in four cases) no observation from before the outburst. 
In the case of Gaia~17bpi (shown in red) we use the date of the brightest $G$-band magnitude, 2019-05-29. However, at the time of writing, there is no indication that the light curve has reached its peak. 

No obvious trend in X-ray luminosity as a function of time since outburst is observed in the sample. Highly X-ray luminous FU Ori stars were observed both around the time of the optical peak ($\Delta t < 3$~years) as well as decades later ($40<\Delta t < 100$~years). There are also non-detections of FU Ori stars spanning nearly the whole range of the post-peak phase. The two observations made before outburst peak (Gaia~17bpi and V2775~Ori) both resulted in non-detections, with fairly low $L_X$ limits ($<$10$^{30}$~erg~s$^{-1}$)  given our assumptions on $N_H$ and $kT$. An observation of HBC~722 at the approximate time of the peak also yielded a non-detection, albeit with a higher upper limit. 

Both of the FU Ori stars that were observed multiple times shortly after the outburst peak (HBC~722 and V960~Mon) increased in luminosity with the subsequent observations. Furthermore, the detection of high-$L_X$ FU Ori stars at late times suggest that an enhancement of X-ray luminosity can be long lasting. However, more data are needed before a definitive conclusion can be made about the time evolution of FU Ori stars in the various stages of their life cycles.

\subsection{X-ray Luminosity Function of FU Ori Stars}\label{coup_comparison.sec}

One method of examining the effect of an FU Ori outburst on X-ray emission from a YSO is through comparison of the X-ray luminosity function (XLF) for FU Ori stars to ordinary YSOs. For this comparison we use the COUP sample, which has a well-studied XLF \citep{2005ApJS..160..379F} that provides a good match to the shape of the distribution of X-ray luminosities of YSOs in other star-forming regions \citep{2008ApJ...675..464W,2015ApJ...802...60K}.

\begin{figure}[t!]
\centering
\includegraphics[width=0.45\textwidth]{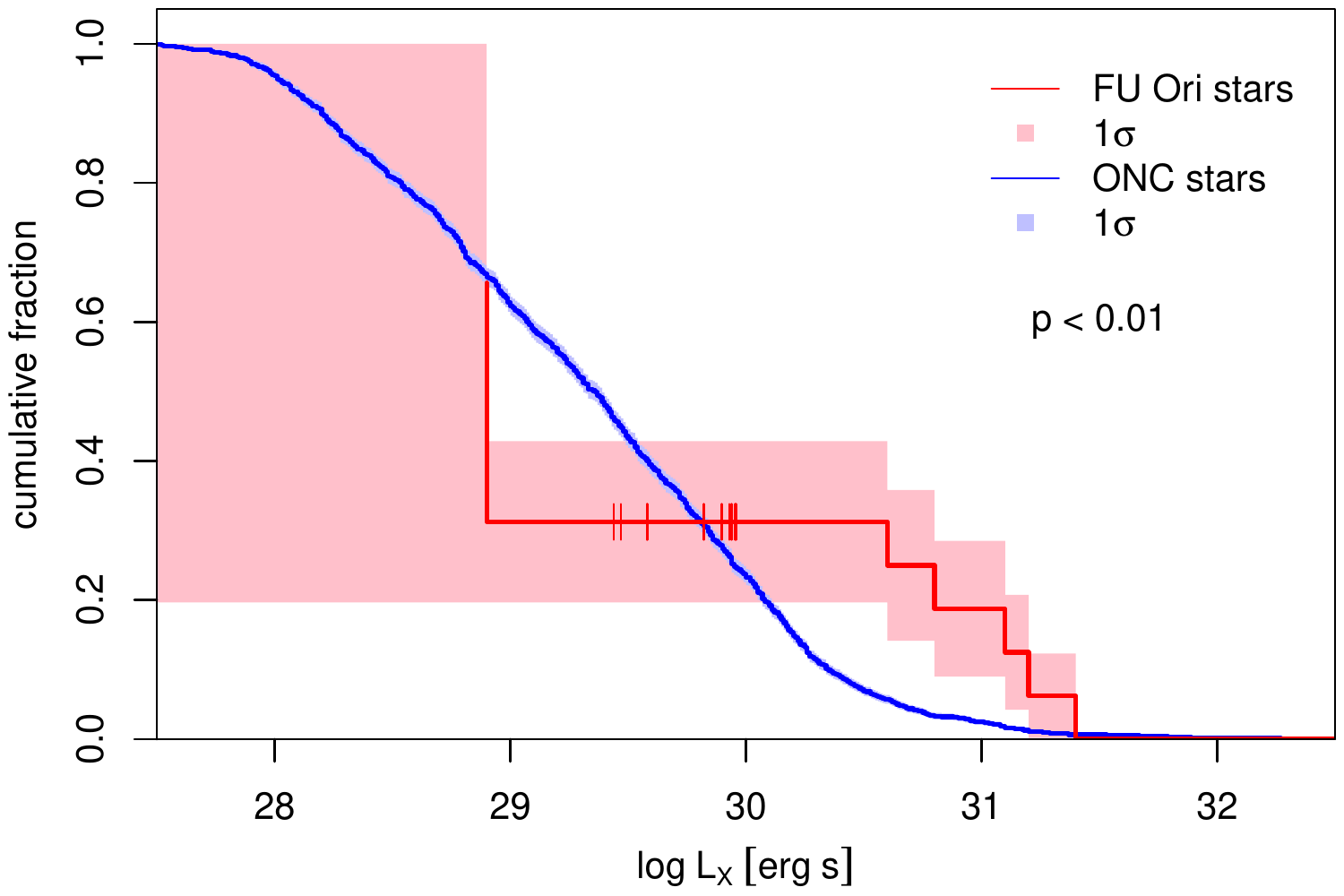} \\
\caption{Cumulative distributions of $L_X$ for FU Ori-type stars (red) and the COUP samples (blue) estimated using the Kaplan--Meier statistic. Tic marks indicate upper limits and steps indicate detections for FU Ori stars. Upper limits derived for COUP stars are included in the derivation of the line, but the tic marks are suppressed on the plot to avoid clutter. The envelopes around both lines indicate uncertainty. 
 \label{kmtest.fig}}
\end{figure}

The Kaplan-Meier statistic \citep{KaplanMeier58} can be used to estimate the cumulative distribution function for a dataset with censoring, in this case censoring of low values of X-ray luminosity as a result of limits on detector sensitivity \citep[see][]{1985ApJ...293..192F}. Figure~\ref{kmtest.fig} shows the Kaplan-Meier estimator for X-ray luminosities -- steps are drawn at the value of detected sources, and tic-marks indicate upper limits on $L_X$, and the shaded region shows the standard deviation of the statistic.   
For objects that have multiple entries in Table~\ref{other.tab}, we base the following analysis on the most recent good measurement of $L_X$.\footnote{This excludes the 2008 epoch for FU Ori showing a possible flare and the non-constraining upper limit from the from the 2014 epoch for V733 Cep. However, these exclusions do not have a significant effect on the results.}
Due to the small number of sources, the uncertainty in the shape of the distribution is large and becomes unconstrained for X-ray luminosities below the least luminous detected source. The Kaplan-Meier estimator is also shown for the sample of COUP X-ray sources (Section~\ref{generic_yso.sec}), with upper limits determined for undetected stars (not shown on the plot) from \citet[][their Table~11]{2005ApJS..160..319G}. Due to the large size of the COUP sample, the standard deviation of the Kaplan-Meier statistic for these data is nearly invisible on the scale of the plot. 

The estimated FU Ori X-ray luminosity function shows that about one third of the FU Ori stars have X-ray luminosities $>$10$^{30.5}$~erg~s$^{-1}$. However, there is a considerable gap, by about an order of magnitude in $L_X$, between the high X-ray luminosity sources and many of the upper limits. The location of the upper limits suggests that the bottom two-thirds of the FU Ori population in $L_X$ have luminosities consistent with those of typical YSOs.

Log-rank tests are typically used to test for differences in distributions with censored values, with different weighting schemes giving rise to different types of tests. We are particularly interested in differences in distributions at high luminosities, so that we can determine whether FU Ori stars tend to be more X-ray luminous than T Tauri stars. However, many of the tests, such as the log-rank test with no weighting, would be more sensitive to the low-luminosity end of the distribution. We chose the $G$-$\rho$ class of tests by \citet{harrington1982}, with $\rho=10$, implemented by the function $\mathtt{survdiff}$ in the {\it R} package {\it survival} \citep{survival-book}. We note that this choice of $\rho$ is somewhat unusual because it puts significant emphasis on the high-luminosity end of the distribution; nevertheless it is still a valid test statistic.\footnote{We are most interested in the high-luminosity tail of the distribution, i.e.\ whether there are significantly more FU Ori stars that have X-ray luminosities above the $\sim$95\% quantile of T Tauri star luminosities. For the Harrington-Fleming test, weights are defined as $w_i(\rho)=\left[ \hat{S}(L_{X,i}) \right]^\rho$, where $\hat{S}$ is the pooled survival probability at the luminosity $L_{X,i}$ of the $i$-th star. For this end of the distribution, typical values of weights would be $w\sim0.95^{10}\sim0.6$, which is reasonable. The test with this choice of $\rho$ would be relatively insensitive to differences in the distributions below the 70\% quantile of $L_X$.} 

Given the small number of FU Ori stars, we calculate the $p$-value for the hypothesis test using simulations of the test statistic, rather than assuming the asymptotic distribution calculated by \citet{harrington1982}. To simulate observations of FU Ori stars (assuming the null hypothesis that the distribution of X-ray luminosities for the FU Ori sample and the COUP sample are drawn from the same underlying distribution), we randomly draw 16 $L_X$ values from the COUP sample and censor them based on the detection limits determined for each observation of the actual FU Ori stars, from which we compute the test statistic. From 10,000 simulations, we are able to reject the null hypothesis with $p<0.01$.

This result can be seen more intuitively using the following, alternative probabilistic argument. From the estimated cumulative distribution of COUP $L_X$ values, it can be seen that only 7\% of objects in the COUP sample have $L_X>10^{30.5}$~erg~s$^{-1}$, while 5 out of 16 FU Ori stars do. If the $L_X$ values for the FU Ori stars are randomly drawn from the same distribution as the COUP stars, the probability of drawing 5 or more stars with $L_X>10^{30.5}$~erg~s$^{-1}$ would be
\begin{equation}
p = \sum_{k=5}^{16}0.07^k0.93^{(16-k)}{16 \choose k} \approx 0.004.
\end{equation}
The results of both tests indicate that the X-ray luminosity function for FU Ori stars differs from that of T Tauri stars, in the sense of FU Ori stars being more luminous. The result using the Harrington-Fleming test remains statistically significant even if only 4 stars were considered to have $L_X > 10^{30.5}$, for example if we used $L_X$ values from the first X-ray observations rather than the most recent, turning HBC~722 into a non-detection. 

For the COUP sample, the top 30\% most luminous stars have $L_X > 10^{29.8}$~erg~s$^{-1}$ based on the Kaplan-Meier estimator for the distribution. Given that the top 30\% FU Ori stars have $L_X > 10^{30.5}$~erg~s$^{-1}$, this suggests that the enhancement of X-ray luminosity for the X-ray luminous FU Ori stars is $\sim$0.7~dex (or roughly a factor of 5).

\subsection{Effects of Mass and YSO Class on $L_X$}\label{mass.sec}

Masses of stars undergoing FU Ori outbursts are poorly constrained due, in most cases, to a lack of pre-outburst spectroscopy. Modeling of post-outburst SEDs, which are dominated by disk emission, provides indirect estimates of stellar properties, suggesting that most of these stars are low-mass \citep{1996ARA&A..34..207H,2014AJ....147..140G}.
 
\begin{figure}[t!]
\centering
\includegraphics[width=0.45\textwidth]{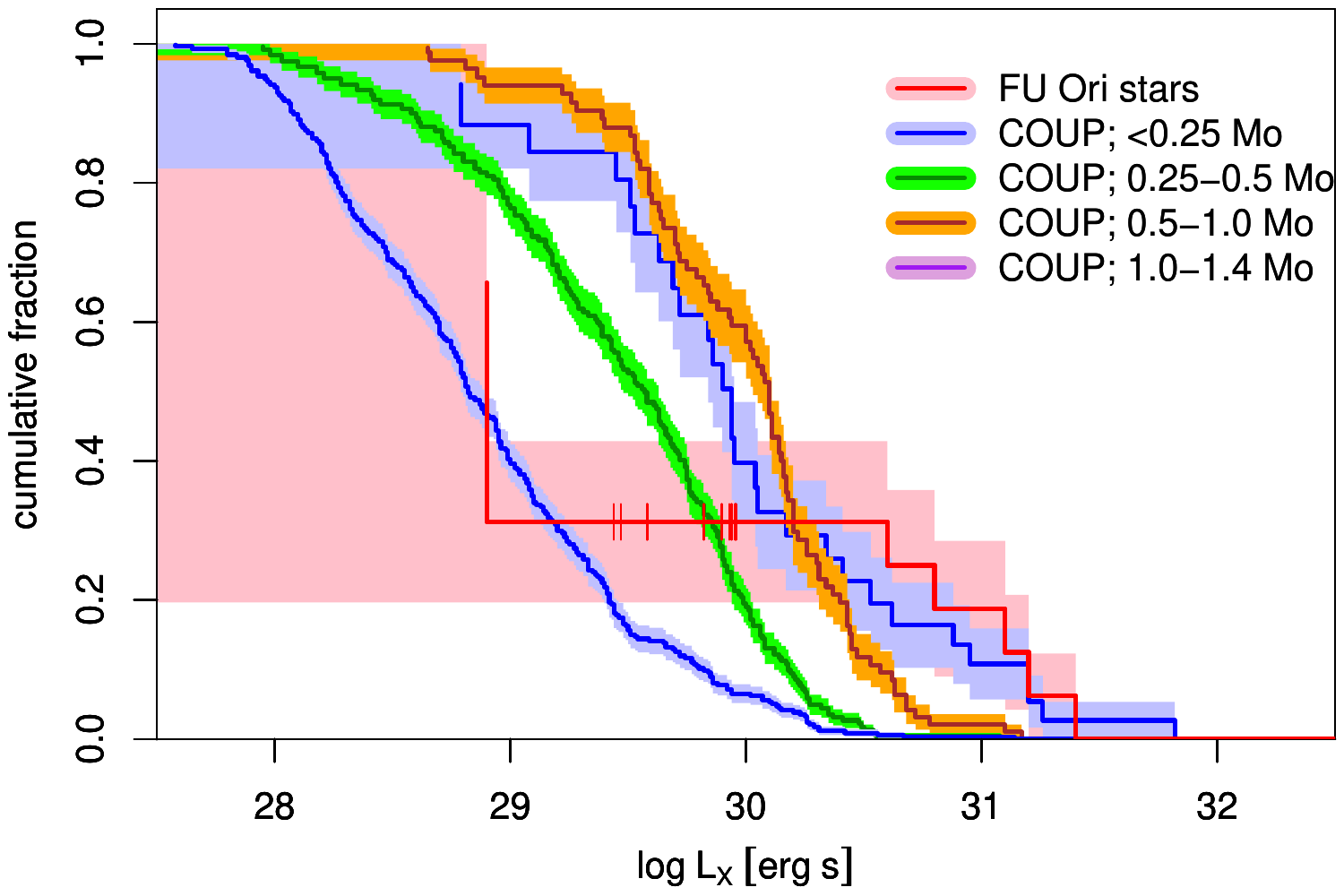} 
\caption{The red line showing the Kaplan-Meier estimator for the FU Ori X-ray luminosity function is that same as in Figure~\ref{kmtest.fig}, but this is compared to COUP stars in restricted mass ranges. 
 \label{kmtest_mass_compare.fig}}
\end{figure}

Figure~\ref{kmtest_mass_compare.fig} shows the XLF for FU Ori stars (full sample) compared to stars in the COUP sample which have been stratified by mass. For COUP stars, we use masses from \citet{2012ApJ...748...14D} derived using the \citet{1998A&A...337..403B} evolutionary models and divide stars into groups with mass ranges $M\leq 0.25$~$M_\odot$, $0.25 < M \leq 0.5$~$M_\odot$, $0.5<M \leq 1.0$~$M_\odot$, and $1.0 < M \leq 1.4$~$M_\odot$. The different mass strata have different luminosity functions, reflecting the well-known $L_X$--mass dependence \citep[][]{2005ApJS..160..401P}.  The two lowest mass strata in the COUP sample contain very few stars with $L_X>10^{30.5}$~erg~s$^{-1}$ so the difference between them and the FU Ori stars is statistically significant ($p<0.001$). Even the $0.5<M\leq 1.0$~$M_\odot$ COUP sample has significantly fewer high X-ray luminosity sources than the FU Ori sample ($p=0.006$). The FU Ori stars most closely follow the distribution for stars with $1.0 < M \leq 1.4$~$M_\odot$ ($p=0.8$). However, it seems unlikely that all 5 X-ray luminous FU Ori stars have masses this high, given that most of them have mass estimates lower than 1~$M_\odot$ \citep{2014AJ....147..140G}. 

FU Ori stars have been classified by their SED shapes as either Class I or Class II YSOs, with a predominance of Class~I objects \citep{2014AJ....147..140G}. This raises the question of how SED stage affects X-ray luminosity and whether this could be the cause of higher $L_X$ seen for FU Ori stars. Out of the 5 FU Ori stars with $L_X>10^{30.5}$~erg~s$^{-1}$, FU Ori, Z CMa, and HBC 722 are likely Class~I, and V1735 Cyg and V960 Mon are likely Class II \citep{2014AJ....147..140G,2015ApJ...801L...5K}. The mixture in classes suggests that class alone does not account for the higher $L_X$. Furthermore, it is thought that X-ray emission is suppressed in both Class~I and Class~II YSOs relative to Class~III YSOs \citep{2007A&A...468..425T,2008ApJ...677..401P}, making it less likely that  the enhancement of $L_X$ for FU Ori stars can be explained by YSO class alone. 

One could ask whether the young stellar objects in the Orion Nebula Cluster are representative of X-ray properties of all young stars in all star-forming environments, and, if there is a difference, how it would affect comparison between our sample of FU Ori stars and generic YSOs.  \citet{2015ApJ...802...60K} examine the X-ray luminosity function of young stellar objects in 17 nearby massive star-forming regions, including the COUP sample as well as samples from regions that contain younger stars, like W40 and NGC 6334, and older stars, like NGC 2362. Overall, the X-ray luminosity functions for the different regions appear to have similar shapes over the range of $L_X=10^{30.5}$--$10^{31.5}$~erg~s$^{-1}$ (their Figure~1), with the exception that some regions with stars older than $\sim$5~Myr have steeper drop-offs with luminosity. None of the regions (including those with large numbers of Class~I objects) appear to have a relative excess of high X-ray luminosity stars compared to Orion. Thus, it is unlikely that the T Tauri stars in any of these regions would more closely resemble the FU Ori X-ray luminosity function. 

\subsection{The Effect of X-ray Flares on $L_X$ Distributions}\label{flare.sec}

The X-ray light curves for the FU Ori stars do not show evidence of large-amplitude flares, with the possible exception of the 2008 epoch for FU Ori \citep{2010ApJ...722.1654S}. Thus, in the previous sections, we compared the FU Ori sample to the sample of low-mass  stars in COUP without significant flare activity. The flaring stars in COUP are among the most X-ray luminous sources in the sample, so inclusion of these sources would broaden that tail of the COUP luminosity function possibly making the difference between the FU Ori XLF and COUP XLF less significant. To investigate this we repeat the Harrington-Fleming test from above, this time including both non-flaring and flaring COUP sources. We find $p=0.02$, which is still marginally significant, but not as strong an effect as when non-flaring sources are compared. 

\section{Discussion and Conclusion}\label{discussion.sec}

We used a sample of FU Ori stars observed by {\it XMM Newton} and {\it Chandra} to investigate whether there is statistical evidence that the outbursts of these stars affect their X-ray emission. The data include a new $\sim$35~ks {\it XMM Newton} observation of the most recent FU Ori outburst, Gaia~17bpi, as well as previously published and archival observations of 15 other FU Ori-type outbursts, yielding a sample (including upper limits for non-detections) that include over half of all known FU Ori stars.

Our analysis shows that the most luminous X-ray sources in our FU Ori sample -- the 5 sources with $L_X>10^{30.5}$~erg~s$^{-1}$ out of 16 objects -- are too luminous to have been drawn by chance from the distribution of $L_X$ for generic YSOs. These stars comprise about 30\% of the FU Ori sample, making them a factor of $\sim$5 times more X-ray luminous than the top 30\% of generic YSOs. This effect seems counterintuitive because other, less extreme cases of heightened accretion generally have the opposite effect of diminishing X-ray emission \citep{2007A&A...468..425T,2008ApJ...677..401P}. However, careful analysis of the full dataset shows that high $L_X$ values noted by earlier studies of individual objects \citep[e.g.,][]{2009ApJ...696..766S} are common for FU Ori stars.

Although the data show a statistically significant effect, they do not yet indicate a cause-and-effect relation. One possibility is that the outbursts are the cause of the high X-ray emission; another is that high X-ray emission could increase the chance of an outburst; and finally both X-ray emission and FU Ori outburst could be connected to a common underlying YSO property.

The high temperatures found from X-ray spectral fitting can be used to rule out one of the simplest explanations for the heightened X-ray emission -- accretion shocks. Gas accreting onto a young star at approximately the free-fall velocity can shock heat plasma to at most several million Kelvin \citep[][their Equation~4]{2016ARA&A..54..135H}, which is much less than the temperatures measured for the five most luminous FU Ori stars in Table~\ref{other.tab}, which range from $\sim$2$\times$10$^{7}$ to more than $10^{8}$~K. 
Instead, such temperatures are similar to those measured for magnetically heated coronae of pre--main-sequence stars. Nevertheless, even moderate $N_H$ can obscure soft X-ray emission (e.g., Figure~\ref{main.fig}), so whether a soft component also exists is left relatively unconstrained.

If the outburst of an FU Ori stars is the cause of heightened X-ray emission, it would mean that the eruption leads to a reconfiguration of the magnetosphere in a way that enhances X-ray emission.  This may come about if the disturbance of a star's magnetosphere by the inward advance of the disk enhances reconnection. Furthermore, X-rays could be generated by strengthened magnetic fields in the disk \citep[][]{2014A&A...570L..11L}. However, there has been little theoretical guidance on how FU Ori outbursts affect X-ray production. 

On the other hand, it is also plausible that FU Ori outbursts could either be triggered directly by the high X-ray emission of a young star or by the strong magnetic activity that gives rise to the X-ray emission. Theoretical models of accretion outbursts suggest that magnetic fields and the ionization structure of the disk play key roles in these outbursts \citep{2010ApJ...713.1134Z,2013ApJ...764..141B,2014ApJ...795...61B}, although these models have not investigated how changes in external sources of ionization would affect outbursts. However, a study of X-ray flares, which are comparable in X-ray luminosity to the most luminous FU Ori stars, suggests that they can have a considerable effect on the structure of the dead zone in disks \citep{2006A&A...455..731I}. In light of this, it is worth investigating whether FU Ori outbursts occur at a higher rate in stars that naturally have higher X-ray luminosities.

Although the enhancement of $L_X$ seen for five of the FU Ori stars is statistically significant, it does not appear to be universal, and there is a fairly large gap of $>$0.5~dex between these 5 stars and the remaining 11 stars -- assuming that these 11 have moderate absorbing column densities. However, the non-detections of some stars cannot be considered evidence for suppression of X-ray emission because we would not expect to detect most ordinary YSOs at the distances of the FU Ori stars given the integration times that were used (e.g., Section~\ref{pre.sec}). Furthermore, even an FU Ori star with $L_X\sim10^{30.5}$~erg~s$^{-1}$ might not be detected in the X-ray if absorption were higher than $\sim$10$^{23}$~cm$^{-2}$. We know that such high absorbing columns are possible because they have been inferred for two detected FU Ori stars: FU Ori and HBC~722. Assuming typical gas-to-dust ratios for T Tauri stars, these gas column densities would correspond to $>$50~mag of extinction in the $V$-band, which would completely obscure an optical source. Given that both these stars are seen in the optical, X-ray absorption does not necessarily correspond to optical extinction. This makes sense in the standard picture of FU Ori stars where optical emission is dominated by the disk, while X-ray emission may originate closer to the star in regions that become obscured by the inward advance of the disk. On the other hand, the changes in star-disk geometry that occur with an eruption do not necessarily result in high $N_H$, since moderate $N_H$ has been observed for systems like V960 Mon and V1735 Cyg.

Several of these scenarios can be tested observationally through follow-up monitoring of FU Ori stars in the X-ray. For example, if outbursts cause increases in X-ray emission, then X-ray properties should be observed to change during outbursts. On the other hand, if the high X-ray luminosity of a YSO statistically increases the outburst rate, these stars should also appear as high $L_X$ objects before the beginning of their outbursts. So far, no known FU Ori star has been observed with {\it XMM Newton} or {\it Chandra} before an eruption. However, with ongoing time-domain surveys searching for FU Ori outbursts, it is increasingly likely that an outburst will be discovered in a field already observed by one of these telescopes. X-ray surveys of nearby star-forming regions ($d < 3$~kpc) have identified tens of thousands of YSOs, and there are likely hundreds of thousands of YSOs projected in these fields that have not been detected \citep{2018ASSL..424..119F}. If the outburst rate were $\sim$10$^{-5}$~yr$^{-1}$, as suggested by \citet{2019MNRAS.486.4590C}, at least one outburst is likely in this sample of stars during the next decade, allowing constraints on pre-outburst X-ray luminosity to be determined. 

\appendix

\section{X-ray Properties of Individual FU Ori Stars}\label{individual.sec}

The X-ray properties or upper limits of FU Ori stars included in Table~\ref{other.tab} are compiled from a variety of sources described below. The ObsIDs for {\it XMM Newton} and {\it Chandra} observations are provided in cases where archival data is reanalyzed.  X-ray luminosities used here are adjusted for X-ray energy band using XSPEC and scaled for assumed distance (Appendix~\ref{distances.sec}).

\subsection{Detected Sources}

\subsubsection{V960 Monocerotis}

\begin{figure*}[h]
\centering
\includegraphics[width=0.45\textwidth]{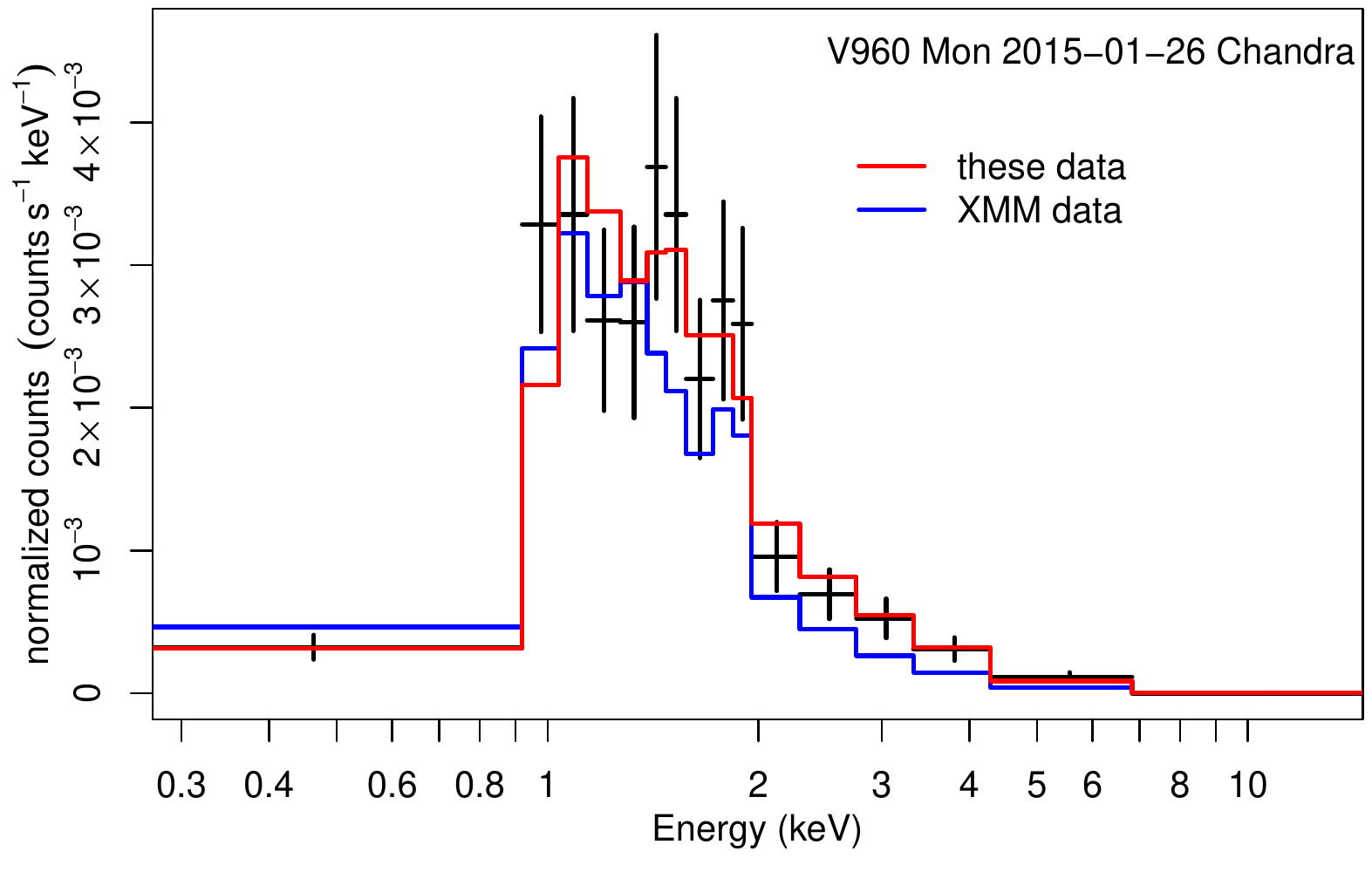} 
\includegraphics[width=0.45\textwidth]{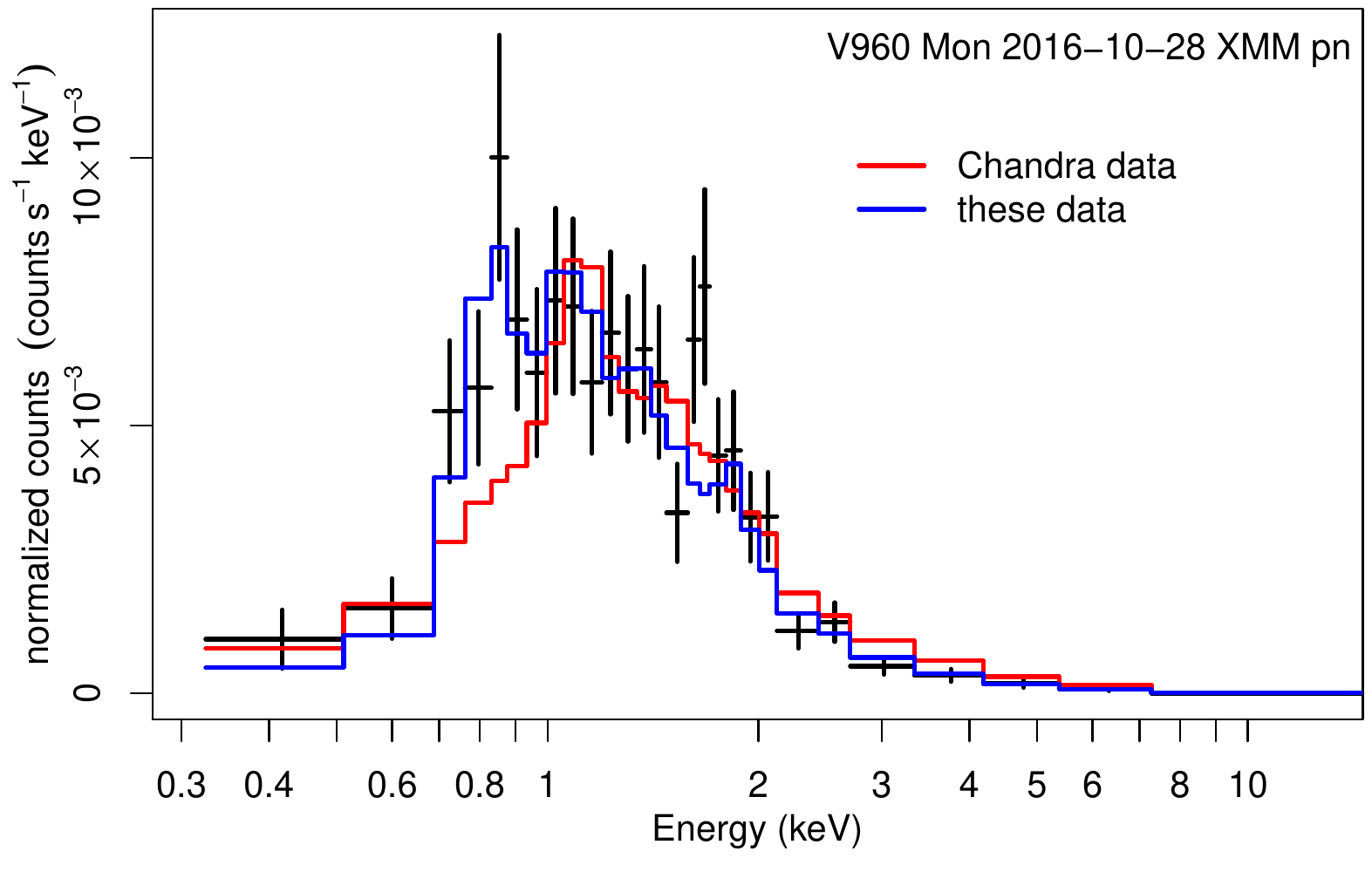} 
\caption{Absorbed plasma models fit to two epochs of X-ray data for V960~Mon observed by {\it Chandra}/ACIS-S (left) and {\it XMM Newton}/EPIC-pn (right). Binned data are shown in black, and the model fits are shown in red ({\it Chandra}) and blue ({\it XMM Newton}). The best-fit parameters indicate that the source is more highly absorbed in the second epoch and that a new $\sim$0.6~keV component has appeared in addition to a $\sim$2.6~keV component. As these changes are difficult to perceive in observed spectra, which are dominated by differences in detector response, we show fits to both epochs overlaid on each plot.  
 \label{xspec_v960mon.fig}}
\end{figure*}

V960 Mon was observed shortly after the beginning of its outburst by {\it Chandra} (ObsID 17587; PI D.\ Pooley) and 1.75 years later by {\it XMM Newton} (ObsID 0781690201; PI M.\ G\"udel). The {\it Chandra} data were taken with the ACIS-S3 chip (49~ks) in ``very faint'' mode. We reduced the data with CIAO \citep{2006SPIE.6270E..1VF}, using the standard data reduction procedures to obtain CCD spectra from imaging data\footnote{\url{http://cxc.harvard.edu/ciao/guides/}}.  The {\it XMM Newton} data was taken using all three EPIC instruments (78~ks) using the medium filter. These data were reduced using the same procedure as in Section~\ref{gaia17bpi_reduction.sec}. For both the {\it Chandra} data and the {\it XMM Newton} data, spectral modeling was performed using XSPEC \citep{1996ASPC..101...17A}. The fits to these spectra are shown in Figure~\ref{xspec_v960mon.fig}. 

For the {\it Chandra} data, we found that a one-temperature absorbed plasma $wabs(apec)$ provided an adequate fit, with $N_H=3.9\pm1.3\times10^{21}$~cm$^{-2}$, $kT=2.6^{+0.8}_{-0.4}$~keV, $\mathrm{norm}=5.6\pm0.8\times10^{-5}$.\footnote{In the {\it apec} model, the the normalization parameter is related to the emission measure, $EM$, by ${\mathrm norm} = \frac{10^{-14} EM}{4\pi D^2}$.} We fixed the elemental abundances in the {\it apec} model to 0.4 times Solar values for consistency with the rest of the analysis and because it is difficult to accurately estimate abundances from CCD spectra. The data were fit to the unbinned spectrum using the $C$-statistic \citep{1979ApJ...228..939C} option in XSPEC, using the {\it steppar} tool to estimate 1$\sigma$ confidence intervals. The parameters obtained this way also provided a good fit to the binned data shown in Figure~\ref{xspec_v960mon.fig}, with a $\chi^2$ value of 8.6 on 12~degrees of freedom. The model fit yields an absorbed X-ray flux of $4.2\times10^{-14}$~erg~s$^{-1}$~cm$^{-2}$ and an absorption-corrected X-ray luminosity of $L_X = 9.4\times10^{30}$~erg~s$^{-1}$ in the 0.5--8.0~keV band. Additional temperature components did not significantly improve the fit. 

 At a separation of 5.6$^{\prime\prime}$ from V960~Mon, a fainter X-ray source (6:59:31.7 $-$4:05:22) can be seen in the {\it Chandra} image that is sufficiently near to be unresolved by {\it XMM Newton}. We fit this source with $N_H=1.9\times10^{22}$~cm$^{-2}$, $kT=1.5$~keV, and $\mathrm{norm}=2.5\times10^{-5}$, yielding an observed flux of $F_X = 6.0\times10^{-15}$~erg~s$^{-1}$~cm$^{-2}$. For the {\it XMM Newton} analysis, this component is included in all models but has a relatively minor effect on the results. 

To model the {\it XMM Newton} spectrum of V960~Mon, we fit both one-temperature  $wabs(apec)$ and two-temperature models $wabs(apec+apec)$ for the main source, along with an additional component for the neighboring source as described above. The data were simultaneously fit to both the pn and MOS detector spectra using XPEC. The one-temperature model for V960~Mon favored a temperature of $kT\approx0.7$~keV; however, it resulted in a moderately poor fit with a  $\chi^2$ value of 75 on 55 degrees of freedom.  For the two-temperature model, $N_H=9.0^{+1.5}_{-2.4}\times10^{21}$~cm$^{-2}$, $kT_1 = 0.54^{+0.09}_{-0.18}$~keV, $\mathrm{norm}_1 = 9.9^{+6.1}_{-6.0}\times10^{-5}$, $kT_2 = 2.6^{+2.9}_{-0.6}$~keV, and $\mathrm{norm}_2 = 2.4^{+1.3}_{-1.2}\times10^{-5}$, yielding an absorbed X-ray flux of $F_X=2.7\times10^{-14}$~erg~s$^{-1}$~cm$^{-2}$ and an absorption-corrected X-ray luminosity of $L_X = 2.3\times10^{31}$~erg~s$^{-1}$ in the 0.5--8.0~keV band. The two-temperature model provides an adequate fit with a $\chi^2$ value of 59 on 53 degrees of freedom. 

Between the first and second epochs several changes in the X-ray spectrum have occurred. The overall absorption has increased by a factor of more than 2. The plasma temperature at the first epoch is 2.6~keV, while at the second epoch one of the plasma components still has a temperature of 2.6~keV, albeit less constrained by the model, with a decrease in emission measure by a factor of $\sim$2. This suggests that the magnetic activity that produced the hotter component of the corona of V960 Mon may have persisted between the observations, while possibly decreasing in strength. However, in the second epoch the analysis strongly indicates the presence of a new, cooler temperature component of $\sim$0.6~keV. The existence of this component is supported by both one- and two-temperature fits. Both 0.6~keV and 2.6~keV components are sufficiently hot that they most likely result from magnetic-reconnection heating of a corona. 

\subsubsection{Z CMa}

Z CMa was observed in 2003 and 2008 by {\it Chandra} \citep{2006A&A...457..223S,2009A&A...499..529S}. The Z CMa system contains a Herbig AeBe star and an FU Ori star, with the former contributing little to no X-ray flux \citep{2009A&A...499..529S}.  \citet{2009A&A...499..529S} state that both epochs are consistent with the same absorbed two-temperature spectral model and report an observed flux of $\sim$1.6$\times10^{-14}$~erg~s$^{-1}$~cm$^{-2}$. Based on their (loosely constrained) best-fit model to the spectrum, we used XSPEC to calculate an absorption corrected X-ray luminosity in the 0.5--8.0~keV band of $L_X = 1.1\times10^{31}$~erg~s$^{-1}$. For the $L_X$ value tabulated in Table~\ref{other.tab}, we do not add in X-ray emission from a nearby source that \citet{2009A&A...499..529S} attribute to a jet from Z CMa.

\subsubsection{HBC~722}

X-ray properties of HBC~722 for the second and third epochs in Table~\ref{other.tab} are based on fits by \citet{2014A&A...570L..11L} to {\it XMM Newton} and {\it Chandra} (unbinned) data. 

\subsubsection{V1735 Cygni}

X-ray properties of V1735 Cyg are obtained from the model fit by \citet{2009ApJ...696..766S} to {\it XMM Newton} data. 

\subsubsection{FU Orionis}

X-ray properties for FU Ori are based on {\it XMM Newton} and {\it Chandra} observations by \citet{2006ApJ...643..995S,2010ApJ...722.1654S}. Both observations show an unusual double-peaked X-ray spectrum, composed of a hard component that is likely associated with the outbursting star and a soft component that is likely associated with a companion \citep{2010ApJ...722.1654S}. For each epoch, we use XSPEC to estimate the luminosity of the hard component alone in the 0.5--8~keV band based on the two-component models from \citet{2006ApJ...643..995S,2010ApJ...722.1654S}. 
The {\it Chandra} observation shows variability that could be a large coronal flare, so we use the  {\it XMM Newton} measurement of $L_X$ as the fiducial ``quiescent'' X-ray luminosity of this object in our analysis. 

\subsubsection{L1551 IRS 5}

L1551 IRS 5 has been observed by {\it XMM Newton} and {\it Chandra} multiple times between 2000 and 2009 \citep{2003ApJ...584..843B, 2011A&A...530A.123S}. \citet{2011A&A...530A.123S} interpret the detection of a marginally extended X-ray source as emission from a jet. They find two possible fits to the X-ray spectrum, with their favored solution having the parameters $N_H=1.1\times10^{22}$~cm$^{-2}$, $kT=0.6$~keV, and $EM = 7.9\times10^{51}$~cm$^{-3}$. This model yields $L_X=10^{28.9}$~erg~s$^{-1}$, which is the lowest $L_X$ for a detected FU Ori star in our sample. An alternate solution, which \citet{2011A&A...530A.123S} argue is less physically plausible, yields an even lower $L_X$ value. However, it is possible that we are not seeing X-ray emission from the star due to high absorption.

\subsection{Non-Detections}\label{appendix_nondetections.sec}

Table~\ref{ulim_others.tab} lists information about upper limits from non-detections, including the observation that was used, the detector for which upper limits are calculated, and the upper limits on flux in the 5 {\it XMM Newton} bands. Notes about upper limits for individual sources (including one derived from {\it Chandra} data) are given below. Unless otherwise stated, all upper limits were calculated based on data from 3XMM-DR7 with the FLIX tool using an aperture of 30$^{\prime\prime}$ and the energy conversion factors from the 3XMM-DR7 documentation\footnote{\url{http://xmmssc.irap.omp.eu/Catalogue/3XMM-DR7/ECF_DR7.txt}}.

\begin{deluxetable}{lrccrrrrr}[t]
\tablecaption{Upper Limits on X-ray Fluxes\label{ulim_others.tab}}
\tabletypesize{\small}\tablewidth{0pt}%\rotate
\tablehead{
  \colhead{Object} &  \colhead{ObsID}   &  \colhead{PI} &\colhead{Detector} & \multicolumn{5}{c}{Upper Limit [$10^{-15}$ erg s$^{-1}$ cm$^{-2}$]}\\
  \colhead{} &  \colhead{}   &  \colhead{} &\colhead{} &  \colhead{Band 1} & \colhead{Band 2} & \colhead{Band 3} & \colhead{Band 4} & \colhead{Band 5}}
\startdata
\object[V883 Ori]{V883 Orionis} &	0205150401 	 & S.\ Skinner & pn 			&2.66	&2.97&	4.34&	16.34&	83.92 \\
~~~--- & 	0205150501		  & S.\ Skinner & pn &2.15			&2.57&4.06&13.21& 62.25\\
\object[V2775 Ori]{V2775 Orionis} & 0503560301	  & S.\ Wolk & MOS1 			&1.75&6.76&12.64&14.19&48.54 \\
\object[V900 Mon]{V900 Monocerotis} & 0783890101	  & D.\ Pizzocaro & pn 		&0.31&0.35&0.74&2.77&18.7  \\
\object[V1515 Cyg]{V1515 Cygni} & 	0402840101	 & S.\ Skinner & pn 			&0.88&1.34&1.41&4.26&21.15 \\
HBC 722 & 0656780701		  & N.\ Schartel & pn 				&1.43&4.04&13.87&25.06&23.08  \\
\object[V2494 Cyg]{V2494 Cygni} &	0720890101	 &  S.\ Wolk & pn 	&0.68&0.80&1.50&5.18&43.04  \\
\object[V1057 Cyg]{V1057 Cygni} & 	 0302640201	 & S.\ Skinner & pn 			&1.16&1.55&1.82&6.16&27.49  \\
\object[V2495 Cyg]{V2495 Cygni} & 	0691580101 	 &  S.\ Wolk & pn &0.57&0.64&1.02&3.69&20.25 \\
~~~~---   &  0720890101 		  &  S.\ Wolk & pn &0.71&0.66&1.19&4.38&34.58  \\
\object[V733 Cep]{V733 Cephei} & 9919,10811,10812	 & T.\ Allen & ACIS-I &\nodata&\nodata&\nodata&\nodata& \nodata \\
~~~~--- & 	0743980301  	&  G.\ Israel & pn & 4.65&4.81&8.69&33.04&225.08\\
\object[IRAS 05450+0019]{IRAS 05450+0019}  & 0201530101	 & S.\ Skinner & pn &1.13&0.99	&1.79&5.95&45.69 \\
\enddata
\end{deluxetable}

To derive the upper limit for the initial {\it XMM Newton} observation of HBC~722, we use a smaller aperture with a radius of 15$^{\prime\prime}$ due to crowding by nearby X-ray sources with overlapping point-spread functions. The FLIX tool identifies flux in several bands, but it is uncertain whether this flux comes from HBC~722 or from its neighbors. For each band, we report the upper limit as the maximum of 1) the upper limit on flux if no sources is detected or 2) a value three standard deviations greater than the measured flux. 

The {\it XMM Newton} observation of V900 Mon is not included in the 3XMM-DR7 catalog, so we follow a data reduction and analysis procedure similar to the one used for Gaia~17bpi in Section~\ref{gaia17bpi_reduction.sec}. The observation was made on 2016-10-10 using the EPIC instrument with the thin filter. The filtered event list has an $\mathtt{ontime}$ of 79915~s. Upper limits on flux are calculated using the procedure described earlier, within a 15$^{\prime\prime}$ aperture at the location of the star (06:57:22.2 $-$08:23:18 ICRS) with the conversion from count rate to flux based on the source ARF.  V900 Mon was also observed by {\it Chandra} in 2007 (ObsID 7475; PI G.\ Garmire), which also resulted in a non-detection. Due to uncertainty in the date of the eruption, it is unknown whether the star was in an FU Ori state during this observation, so we do not include it in our analysis. 

The source V733 Cep was observed in three observations by {\it Chandra}, which were included in the Star Formation in Nearby Clouds \citep[SFiNCs;][]{2017ApJS..229...28G} catalog. SFiNCs provides deep X-ray source lists for this region generated by stacking the all three observations and using sensitive source extraction techniques; however, an X-ray counterpart to V733 Cep was not detected. We empirically estimate the completeness limit of the SFiNCs catalog for sources within 5$^\prime$ of V733 Cep that have a $N_H$ between $10^{21.5}$ and $10^{22}$~cm$^{-2}$. The flux distribution for these sources peaks at $F_{X}=10^{-13.9}$~erg~s$^{-1}$~cm$^{-2}$ in the 0.5--8.0 keV band, which we interpret as the upper limit for V733 Cep. V733 Cep was also included in an {\it XMM Newton} observation (ObsID 0743980301; PI G.\ Israel), from which we calculated upper limits based on the $pn$ detector. Given that this observation provides much weaker constraints than the stacked {\it Chandra} observations, we use the {\it Chandra} upper limit for the analysis in this paper. 

Upper limits for V1057 Cyg and V1515 Cyg were previously calculated by \citet{2009ApJ...696..766S} using the same data sets as us. In the former case we find a similar upper limit, while in the latter case our upper limit, using a different assumed distance and spectral model, is lower.

\section{Distances to FU Ori Stars}\label{distances.sec}

The {\it Gaia} DR2 catalog \citep{2016A&A...595A...1G,GaiaBrown} provides the opportunity to refine distances to various FU Ori stars. Although a few FU Ori stars have reliable parallax measurements, with small formal uncertainties and renormalized unit weight errors that are less than 1.4 as recommended by the {\it Gaia} team\footnote{\url{https://www.cosmos.esa.int/web/gaia/dr2-known-issues}}, many of the stars do not. Thus, we derive new {\it Gaia}-based distances to the stellar associations that host these objects and assume that the FU Ori stars lie at the same distance.

Our procedure is similar to that described by \citet{2018ApJ...869..146H}; we obtain a catalog of probable association members from the literature, match these stars to {\it Gaia} DR2 sources, discard stars with parallaxes that are discrepant by $>$3$\sigma$, and find the median parallax of the resulting sample. 

Several groups have attempted to estimate the zero-point offsets in the {\it Gaia} DR2 parallaxes. 
For the list of star-forming regions below, we report median parallaxes in the {\it Gaia} DR2 system but apply a zero-point correction of 0.0523~mas \citep{2019arXiv190208634L} when calculating distances, making them slightly smaller. In all cases, uncertainties on the median parallaxes are dominated by the $\sim$0.04~mas spatially correlated systematic uncertainty reported by the \citet{GaiaBrown}. 

\begin{description}
\item[G53.2]  Gaia~17bpi is a probable member of the G53.2 region. The median parallax to probable members of this region is found to be 0.79~mas \citep{2018ApJ...869..146H}, from which we calculate a distance of $1200^{+80}_{-70}$~pc.
\item[CMa~R1] V900~Mon, V960~Mon, and Z~CMa are probable members of CMa~R1. We estimate a median parallax of 0.84~mas for stars in the association based on 88 probable early-type members cataloged by \citet{1999MNRAS.310..210S}, yielding a distance of 1120~pc. This is consistent with the classical calculation of 1150~pc by \citet{1974AJ.....79.1022C}. 
\item[Cygnus X] V1515 Cyg is located in NGC 6914, a star-forming region projected on the sky near Cygnus~X. V1515 Cyg has a reliable {\it Gaia} DR2 parallax of 0.99$\pm$0.03, from which we obtain a distance of 960~pc.
\item[Cep~OB3] V733 Cep is a probable member of the Cep~OB3 association, which has a median parallax of 1.16 \citep{2019ApJ...870...32K}, corresponding to a distance of 825~pc.
\item[NGC~7000] V1057~Cyg and HBC~722 are both probable members of NGC~7000. We find a median parallax of 1.21~mas for  member candidates from the catalog of \citet{2017A&A...602A.115D}, corresponding to 795~pc.
\item[IC 5146] V1735 Cyg is associated with IC 5146, which has a median parallax of 1.28 \citep{2019ApJ...870...32K}, corresponding to a distance of 752~pc. 
\item[Cyg OB7] V2494 Cyg and V2495 Cyg are both probable members of  Cyg OB7. We find a median parallax of 1.63~mas for  member candidates from \citet{2006AJ....131.1530H}, corresponding to 594~pc.
\item[Orion Complex] V883 Ori, FU Ori, V2775 Ori, and IRAS 05450+0019 are located in the Orion complex. V883 Ori is located in the head of the Orion~A cloud, south of the Orion Nebula Cluster, in a region where stars have an average parallax of 2.50~mas \citep{2018A&A...619A.106G}, corresponding to a distance of 392~pc. FU Ori is located in the $\lambda$ Ori region, which has a {\it Gaia}-based distance of 404~pc from \citet{2018AJ....156...84K}. V2775 Orionis is located in the L1641 region, near the tail of Orion~A, in a region with an average parallax of 2.35~mas \citet{2018A&A...619A.106G}, corresponding to a distance 416~pc. IRAS 05450+0019 is located in NGC 2071. The median parallax is 2.19~mas, based on a sample of YSO candidates from \citet{2012AJ....144..192M} that lie within 15$^\prime$ of IRAS 05450+0019, corresponding to a distance of 446~pc.
\item[Taurus] L1551 IRS 5 is a highly absorbed protostar in the Taurus association that is undetectable by {\it Gaia}. The distance derived by \citet{2018AJ....156..271L} for L1551 is 145~pc. After applying the \citet{2019arXiv190208634L} zero-point correction, this becomes 144~pc. 
\end{description}

\section{Optical and ultraviolet photometry}\label{om.sec}

\begin{figure}[t]
\centering
\includegraphics[width=0.45\textwidth]{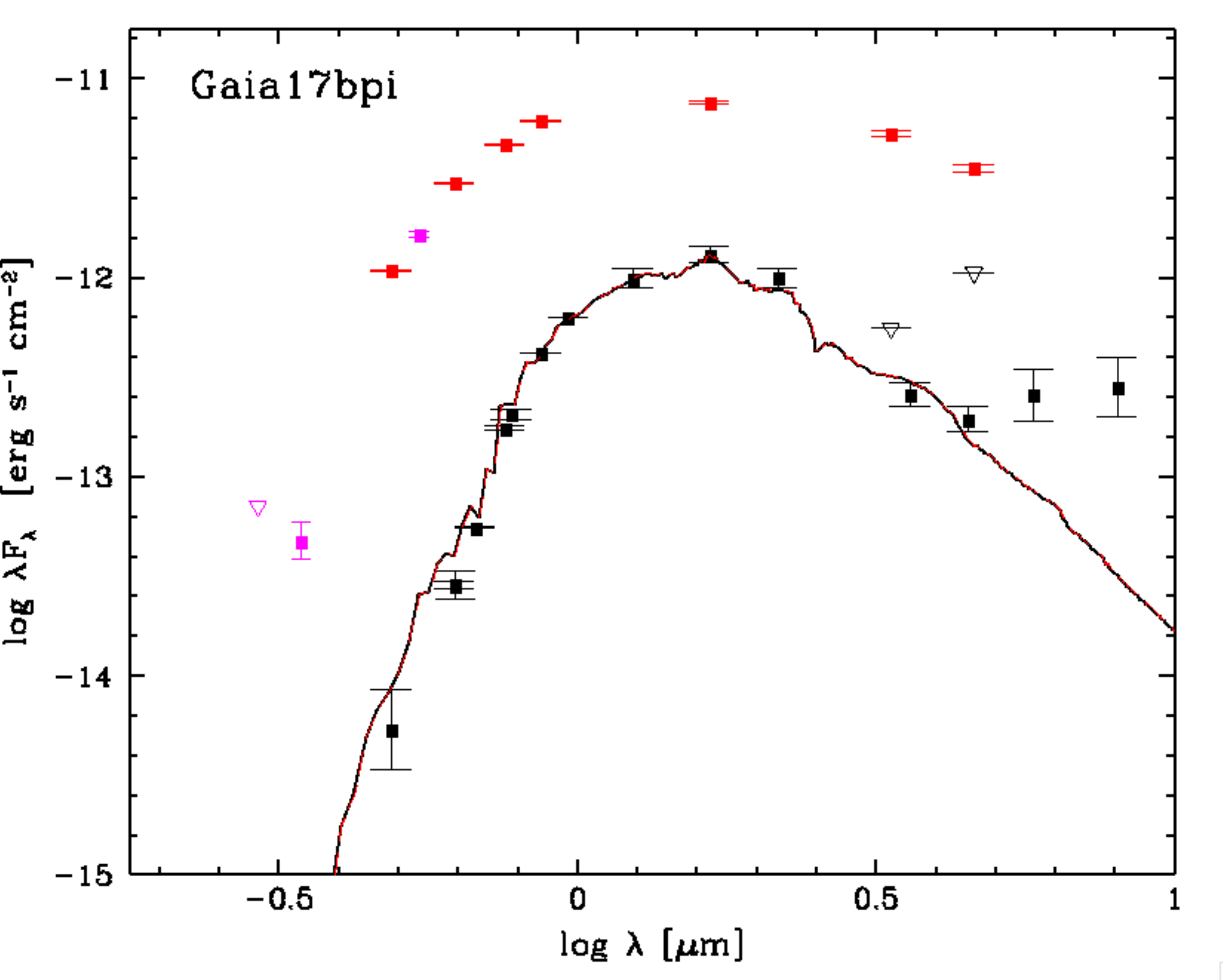} 
\caption{SED of Gaia~17bpi pre-burst (black points and line) and during the outburst (red and magenta points). The magenta points show the new $U$ and $V$ flux measurements and the $UVW1$ upper limit from the OM. Other points are from \citet{2018ApJ...869..146H}.
 \label{sed.fig}}
\end{figure}

Figure~\ref{sed.fig} shows the SED of Gaia~17bpi during its eruption using photometry from \citet{2018ApJ...869..146H} with the {\it XMM Newton} OM $UVW1$, $U$, and $V$-band photometry newly added here. The sources is clearly more luminous in $U$ and $V$ than it would have been in its pre-outburst state (shown for comparison). 

\acknowledgements We thank the {\it XMM Newton team} for approving Gaia~17bpi as an unanticipated TOO. We would like to thank Hannah Earnshaw for advice on reducing the {\it XMM Newton} data, Matthew Povich for useful discussions about star-forming regions, and the anonymous referee for suggestions that improved the manuscript.  

\facility{{\it XMM Newton}, {\it Chandra X-ray Observatory}} 

\software{
          CIAO \citep{2006SPIE.6270E..1VF},
          HEASOFT \citep{2014ascl.soft08004N},
          PIMMS \citep{1993Legac...3...21M},
          R \citep{RCoreTeam2018}, 
          SAS \citep{2014ascl.soft04004S},
          survival \citep{survival-book},
          TOPCAT \citep{2005ASPC..347...29T},
          XSPEC \citep{1996ASPC..101...17A}
          }

\bibliography{main_arxiv.bbl}

\end{document}